\documentclass[aps,12pt,final,notitlepage,oneside,onecolumn,nobibnotes,nofootinbib,superscriptaddress,noshowpacs,centertags]{revtex4-2}
\usepackage{graphics}
\usepackage{epsfig,array}
\usepackage{color}
\newcommand{\be}{\begin{eqnarray}}
\newcommand{\ee}{\end{eqnarray}}

\newcommand{\jpsi}{J/\Psi}
\begin{document}

\title{The Covariant Approach for the Radiative Decay of the $\jpsi$ Meson
and Relation to Amplitudes Extracted in the Helicity Basis}

\author{A.~V.~Anisovich}
\affiliation{National Research Center "Kurchatov Institute" -- PNPI,
Gatchina, Russia}

\author{K.~V.~Nikonov}
\affiliation{National Research Center "Kurchatov Institute" -- PNPI,
Gatchina, Russia}

\author{A.~V.~Sarantsev}
          \email{E-mail: sarantsev$\_$AV@pnpi.nrcki.ru}
\affiliation{National Research Center "Kurchatov Institute" -- PNPI,
Gatchina, Russia}

\begin{abstract}
We have developed the covariant approach for the partial wave
analysis of the radiative decay of the vector meson into two
pseudoscalar mesons based on the  orbital momentum- intrinsic spin
decomposition. This allowed us to include the data on the radiative
$\jpsi$ decay into a large data base for analysis of the meson
production reactions. The correspondence between partial wave
amplitudes extracted in our approach and those extracted in the
helicity basis is determined.
\end{abstract}

\date{Received December 5, 2023; revised December 5, 2023; accepted
January 9, 2024}


\maketitle

\section{Introduction}

The search for glueballs is one of the most challenging tasks in the
physics of the strong interactions. The glueball states were
predicted in many theoretical models and lattice calculations. In
all these approaches the scalar glueball is expected to have the
mass below 2 GeV and this energy range is available for many
experiments. Thus the scalar glueball was searched in the $\pi N$
collision reactions, in the proton-antiproton annihilation
processes, in central production reactions and in decays of heavy
particles. Although quite a number of the candidate were suggested,
it was realized that the scalar glueball is not produced as a well
isolated state but it mixes with nearby $q\bar q$ states leading to
a complicated picture.

The radiative decay of the $J/\Psi$ and $\Psi'$ particles is one of
the most prominent reactions for the search of glueballs. After
emission of photon the $c \bar c$ quarks annihilate into two gluons
and therefore one can expect that glueball-reach states are strongly
produced in this reactions.

The BES3 collaboration reported the measurements of the radiative
decay of the $J/\Psi$-meson in a number of final states, in
particular, into channels with 2 pseudoscalar mesons: $\pi\pi$
~\cite{Ablikim:2015umt}, $K\bar K$ \cite{Ablikim:2018izx},
$\eta\eta$ \cite{Ablikim:2013hq}. The partial wave analysis
performed in the helicity basis showed a prominent structure for the
S-wave below 2 GeV and an enhancement in the mass region 2.3 GeV in
the tensor wave. However the observed structure is reveal a
complicated picture with sets of peaks and dips in the extracted
S-wave.

\section{The covariant approach}

The covariant approach for the decomposition of amplitudes in the
intrinsic spin-orbital momentum basis was suggested in a number of
works
\cite{Zemach:1965ycj,Chung:1997jn,Anisovich:1992rd,Zou:2002ar}. The
full mathematical tool for this method was presented in
\cite{Anisovich:2001ra} for the production of bosons and in
\cite{Anisovich:2004zz} for the production of fermion states.  This
method is used in a large set of the partial wave analyses of the
experimental data and is the main approach used by a number of
international collaborations such as Crystal Ball, CB-LEAR, CB-ELSA,
CLAS, HADES and partly BES3. Below we develop this method for the
radiative decay of the vector-meson into two pseudoscalar particles.

Let us consider the radiative decay of the $\jpsi$-meson into two
pseudoscalar mesons (e.g. two pions) where $k_1$ is the momentum of
photon and $k_2$ and $k_3$ are momenta of the final pions. The
reaction can be considered as two particle scattering and the total
amplitude can be decomposed to a sum of the partial wave amplitudes
with the following general structure:
\be
A=\epsilon_\mu^{\Psi*}\epsilon_\nu^{\gamma}
\sum\limits_{SLJ}a_{SLJ}(s)P_{\mu\nu\alpha_1\ldots\alpha_J}^{SLJ}
O^{\alpha_1\ldots\alpha_J}_{\beta_1\ldots\beta_J}D_{\beta_1\ldots\beta_J},
\ee
where $\epsilon^{\Psi}$ is the polarization vector of the initial
$\jpsi$ meson and $\epsilon^{\gamma}$ is the polarization vector of
the final photon. The $J$ describes to the total spin of the two
meson partial wave and $L,S$ corresponds to the orbital momentum and
intrinsic spin between $\jpsi$ and photon. The $a_{SLJ}(s)$ is the
energy dependent part of the partial wave amplitude $s=P^2$ where
$P$ is the momentum of two meson system $P=(k_2+k_3)$. The tensor
$P^{LSJ}_{\mu\nu\alpha_1\ldots\alpha_J}$ describes the production of
the two meson system with total spin $J$, the tensor
$O^{\alpha_1\ldots\alpha_J}_{\beta_1\ldots\beta_J}$ describes the
structure of the system's propagator and the tensor
$D_{\beta_1\ldots\beta_J}$ describes the decay of the system into
two final mesons.

The decay (or transition) of a partial wave into two pseudoscalar
mesons has intrinsic spin $S=0$ and therefore the decay is fully
described by the orbital momentum operator
\be
D_{\beta_1\ldots\beta_J}=X^{(J)}_{\beta_1,\ldots,\beta_J}(k^\perp),
\ee
where
\be
k^\perp_\mu=\frac 12(k_2-k_3)_\nu g^\perp_{\mu\nu},\qquad
g^\perp_{\mu\nu}=\left(g_{\mu\nu}-\frac{P_\mu P_\nu}{P^2}\right
),\qquad k_\perp^2=k^\perp_\mu k^\perp_\mu=(k^\perp_0)^2-({\bf
k}^\perp)^2.
\ee
Here and later the convolution of any two indices always means the
covariant convolution. The expression for the orbital momentum
operators obtained with a recurrent formulas is given in details in
\cite{Anisovich:2004zz}. For the lowest orbital momentum tensors
have the following form:
\be
X^{(0)}=1,\qquad X^{(1)}_\mu=k^\perp_\mu,\qquad
X^{(2)}_{\mu\nu}(k^\perp)=\frac 32\left (k^\perp_\mu
k^\perp_\nu-\frac 13 k_\perp^2 g^\perp_{\mu\nu}\right).
\ee
The highest states can be calculated from the recurrent formulae:
\be
X^{(L)}_{\mu_1\ldots\mu_L}= \frac{2L\!-\!1}{L^2}\Big (
\sum^L_{i=1}X^{{(L-1)}}_{\mu_1\ldots\mu_{i-1}\mu_{i+1}\ldots\mu_L}
k^\perp_{\mu_i}- \frac{2k^2_\perp}{2L\!-\!1}  \sum^L_{i,j=1 \atop
i<j} g^\perp_{\mu_i\mu_j}
X^{{(L-2)}}_{\mu_1\ldots\mu_{i-1}\mu_{i+1}\ldots\mu_{j-1}\mu_{j+1}
\ldots\mu_L} \Big ).
\label{rec}
\ee
The tensor $O^{\alpha_1\ldots\alpha_J}_{\beta_1\ldots\beta_J}$ which
corresponds to the convolution of the polarization tensors of the
state with the spin $J$ is constructed from the metrical tensors
orthogonal to the momentum of the two meson system. For the lowest
spin states it has the form:
\be
O_{\mu}^{\alpha}=g^\perp_{\mu\alpha}, \qquad
O_{\mu\nu}^{\alpha\beta}=\frac
12\left(g^\perp_{\mu\alpha}g^\perp_{\nu\beta}+g^\perp_{\nu\alpha}g^\perp_{\mu\beta}\right
) -\frac 13 g^\perp_{\mu\nu}g^\perp_{\alpha\beta}.
\ee
For the highest spin states this tensor can be calculated with the
recurrent formulae given in \cite{Anisovich:2004zz}. This tensor is
also called "the projection operator" because it projects any tensor
to a tensor with properties which corresponds to the partial wave
with spin $J=L$: orthogonality to the total momentum $P$ and the
traceless property. For example:
\be
k_{\mu_1}\ldots k_{\mu_L} O^{\mu_1\ldots\mu_L}_{\nu_1\ldots
\nu_L}=\frac{1}{\alpha_L}X^{(L)}_{\nu_1\ldots\nu_L}(k^\perp) \ ,
\qquad \alpha_L=\prod^L_{l=1}\frac{2l-1}{l}.
\label{19}
\ee
The orbital momentum tensor is already constructed correspondingly
to the partial wave properties and therefore the projection operator
does not change its structure:
\be
X^{(L)}_{\nu_1\ldots\nu_L}(k^\perp)=
O^{\mu_1\ldots\mu_L}_{\nu_1\ldots \nu_L}
X^{(L)}_{\mu_1\ldots\mu_L}(k^\perp).
\ee

The production tensor depends on the total spin of the two meson
system, the orbital momentum between initial $\jpsi$-meson and
photon and intrinsic spin. The two particles with spin 1 can create
an intrinsic spin $S=1,2,3$ and therefore in the $S$-wave only
states with $J^P=0^+,1^+,2^+$ can be formed. The correspondents
states can be described as:
\be
S^{(0)}=(\epsilon^{\Psi*}\epsilon^{\gamma}),\qquad
S^{(1)}_\mu=\varepsilon_{\mu\nu\alpha\beta}\epsilon^{\Psi*}_\nu\epsilon^{\gamma}_\alpha
P_\beta\equiv\varepsilon_{\mu\nu\alpha
P}\epsilon^{\Psi*}_\nu\epsilon^{\gamma}_\alpha,\qquad
S^{(2)}_{\mu\nu}=\epsilon^{\Psi*}_\alpha\epsilon^{\gamma}_\beta
O^{\alpha\beta}_{\mu\nu},
\ee
where $\varepsilon_{\mu\nu\alpha\beta}$ is the fully antisymmetrical
tensor.

The production tensors for the partial waves with total spin $J$ and
parity $P=(-1)^J$ can be formed as:
\be
&P_{\mu_1\ldots\mu_J}(0,L\!=\!J)&=S^{(0)}X^{(L)}_{\mu_1\ldots\mu_J}(k_1^\perp),
\label{L0}\\
&P_{\mu_1\ldots\mu_J}(1,L\!=\!J)&=\varepsilon_{\mu_1\alpha\nu P}
S^{(1)}_\alpha
X^{(L)}_{\nu\mu_2\ldots\mu_J}(k_1^\perp)\label{L1},\\
&P_{\mu_1\ldots\mu_J}(2,L\!=\!J\!-\!2)&=S^{(2)}_{\mu_1\mu_2}
X^{(L)}_{\mu_3\ldots\mu_J}(k_1^\perp)\label{L2m},\\
&P_{\mu_1\ldots\mu_J}(2,L\!=\!J)&=S^{(2)}_{\mu_1\alpha}
X^{(L)}_{\alpha\mu_2\ldots\mu_J}(k_1^\perp)\label{L20},\\
&P_{\mu_1\ldots\mu_J}(2,L\!=\!J\!+\!2)&=S^{(2)}_{\alpha\beta}
X^{(L)}_{\alpha\beta\mu_1\ldots\mu_J}(k_1^\perp)\label{L2p}.
\ee
However in the case of the real photon, due to gauge invariance only
amplitudes with production tensors (\ref{L0})--(\ref{L2m}) will be
independent.

The partial wave with parity $P=(-1)^{(J+1)}$ can be formed as:
\be
&P_{\mu_1\ldots\mu_J}(1,L\!=\!J-1)&=S^{(1)}_{\mu_1}
X^{(L)}_{\mu_2\ldots\mu_J}(k_1^\perp),\label{L1m}\\
&P_{\mu_1\ldots\mu_J}(1,L\!=\!J+1)&=S^{(1)}_\alpha
X^{(L)}_{\alpha\mu_1\ldots\mu_J}(k_1^\perp),\label{L1p}\\
&P_{\mu_1\ldots\mu_J}(2,L\!=\!J\!-\!1)&=\varepsilon_{\mu_1\alpha\beta
P}S^{(2)}_{\alpha\mu_2}
X^{(L)}_{\beta\mu_3\ldots\mu_J}(k_1^\perp),\label{L2em}\\
&P_{\mu_1\ldots\mu_J}(2,L\!=\!J\!+\!1)&=\varepsilon_{\mu_1\alpha\beta
P}S^{(2)}_{\alpha\nu}
X^{(L)}_{\beta\nu\mu_2\ldots\mu_J}(k_1^\perp).\label{L2pm}
\ee
In the case of the real photon only amplitudes with production
vertices (\ref{L1m}) and (\ref{L2em}) are linear independent.

In the $\jpsi$ radiative decay only isoscalar partial waves are
expected to be produced. Moreover, in the case of the final two
pseudoscalar meson states only partial waves with even spin $J=2n$
and positive parity are contributed to the reaction. For a
convenience we will list these operators:
\be
A_J(0L)&=S^{(0)}X^{(J)}_{\mu_1\ldots\mu_J}(k_1^\perp)O^{\mu_1\ldots\mu_J}_{\nu_1\ldots\nu_J}
X^{(J)}_{\nu_1\ldots\nu_J}(k^\perp)a_{0\,L\,J}(s), &L\!=\!J,\\
A_J(1L)&=\varepsilon_{\mu_1\alpha\beta P} S^{(1)}_\alpha
X^{(L)}_{\beta\mu_2\ldots\mu_J}(k_1^\perp)O^{\mu_1\ldots\mu_J}_{\nu_1\ldots\nu_J}
X^{(J)}_{\nu_1\ldots\nu_J}(k^\perp)a_{1\,L\,J}(s), &L\!=\!J,\\
A_J(2L)&=S^{(2)}_{\mu_1\mu_2}X^{(J-2)}_{\mu_3\ldots\mu_J}(k_1^\perp)
O^{\mu_1\ldots\mu_J}_{\nu_1\ldots\nu_J}
X^{(J)}_{\nu_1\ldots\nu_J}(k^\perp)a_{2\,L\,J}(s), & L\!=\!J\!-\!2.
\label{ampl}
\ee

\section{The energy independent partial wave analysis based on the
helicity approach}

The data on the reaction $J/\Psi \to \gamma \pi^0\pi^0$ and $\jpsi
\to K^+K^-$ were analyzed in the energy independent approach based
on the helicity formalism \cite{Ablikim:2015umt}. Here the data were
divided into the 18 MeV bins in two meson invariant mass and the
obtained angular distributions were analyzed to extract the
contribution from the $S$- and $D$-waves. It was shown that the
$J^P=4^+$ states contributed very little to the data and can be
neglected in the analysis. Let us note that due to absence of the
polarization information it would be very difficult task to perform
such analysis in the presence of the notable contribution from $4^+$
states. The restriction of the analysis by the $S$- and $D$-waves
allowed authors to resolve discrete ambiguity in the partial wave
decomposition  and obtain for every energy two solutions. Both
solutions reproduce exactly the angular dependence at a fixed energy
although one solution produced an unphysical energy dependence, for
example, generating jumps. In this approach the amplitude where
polarization vector of the $\jpsi$ has helicity $M$ and polarization
vector of the photon has helicity $\lambda_\gamma$ can be written as
(see \cite{Ablikim:2015umt}):
\be
U^{M,\lambda_\gamma}({\bf p},s)=\sum\limits_{j,J_\gamma}
V_{j,J_\gamma}A^{M,\lambda_\gamma}_{j,J_\gamma~.}
\ee
Here the $j$ is the total spin of the partial wave and $J_\gamma$
lists the multipoles. Usually these multipoles are called:
\be
E_0=V_{0,1},\qquad E_1=V_{2,1}, \qquad M_2=V_{2,2}, \qquad
E_3=V_{2,3}\,.
\label{mult}
\ee
The angular dependence of the helicity amplitudes is given as:
\be
A^{M,\lambda_\gamma}_{j,J_\gamma}&=&\sum\limits_\mu
c_{j,\mu}^{J_\gamma,\lambda_\gamma} N_{J_\gamma} N_j
\exp^{-iM(\pi+\phi_\gamma)}
d^1_{M,\mu-\lambda_\gamma}(\pi-\theta_\gamma)\exp^{-i\mu\phi_\pi}d^j_{\mu,0}(\theta_\pi)\frac{1}{\sqrt
2}\times\nonumber
\\&&\left[\delta_{\lambda_\gamma,1}+\delta_{\lambda_\gamma,-1}(-1)^{J_\gamma-1}\right
].
\label{hel_ampl}
\ee
The sign of the $\mu$ is the same as $\lambda_\gamma$. For $j=2$ and
$\lambda_\gamma=1$ the $\mu=1,2,3$. The $N_j=\sqrt{(2j+1)/4\pi}$ and
the constants $c_{j,\mu}^{J_\gamma,\lambda_\gamma}$ contain the
Clebsch--Gordan coefficients. For the scalar states all
$c^{j,\lambda}_{00}=1$. For the tensor states:
\be
c_{2,0}^{1,\pm 1}=\sqrt{\frac{1}{10}}\,, &\qquad c_{2,0}^{2,\pm
1}=\pm
\sqrt{\frac{3}{10}}\,, &\qquad c_{2,0}^{3,\pm 1}=\sqrt{\frac{6}{35}}\,,\nonumber \\
c_{2,1}^{1,\pm 1}=\sqrt{\frac{3}{10}}, &\qquad c_{2,1}^{2,\pm 1}=\pm
\sqrt{\frac{1}{10}}\,, &\qquad c_{2,0}^{3,\pm 1}=-\sqrt{\frac{8}{35}}\,,\nonumber \\
c_{2,2}^{1,\pm 1}=\sqrt{\frac{3}{5}}\,, &\qquad c_{2,2}^{2,\pm
1}=\mp \sqrt{\frac{1}{5}}\,, &\qquad c_{2,0}^{3,\pm
1}=\sqrt{\frac{1}{35}}\,.
\ee

\section{Correspondence of the The energy independent partial wave analysis based on the
helicity approach}

The multipoles (\ref{mult}) which were extracted in
\cite{Ablikim:2015umt} and \cite{Ablikim:2018izx} are very important
for the search of the signals from the glueball. In particular the
$S$-wave showed an interesting structure with a set of peaks and
dips. It means that we do not observe a single signal which can be
associated with glueball but a strong interference of the glueball
with standard quark--antiquark states. It means that these data
should be analyzed together with other data measured in the $\pi N$
collision, in meson production in proton--antiproton annihilation
reaction and in the decay of heavy mesons. The combined analysis of
these reactions was carried out in the $SL$ basis and it is
important to be able to include these data in the similar way. Below
we consider the correspondence between the helicity formalism and
our standard approach.

For the $S$-wave the $LS$ amplitude is:
\be
A^{M\Lambda}_0(SL)=A^{M\Lambda}_0(00)=(\epsilon^{\Psi
M*}\epsilon^{\gamma\Lambda}) a_{0}(s)\,.
\label{a0}
\ee
The polarization vectors of $\jpsi$ and photon can be taken in the
same helicity basis (see below) and for convenience we omit $LJ$
indices in the amplitude: $a_0(s)\equiv a_{000}(s)$ and
$\Lambda\equiv \lambda_\gamma$.

In the case of the production of the tensor particle we have three
amplitudes:
\be
A^{M\Lambda}_2(SL)=A^{M\Lambda}_2(20)&=&\epsilon^{\Psi
M*}_\mu\epsilon^{\gamma\Lambda}_\nu O_{\mu\nu}^{\alpha\beta}
X^{(2)}_{\alpha\beta}(k^\perp) a_{20}(s)\,,
\label{a20}
\ee
where $a_{20}(s)\equiv a_{202}(s)$ is the energy dependent part for
the production of the tensor state with intrinsic spin 2.

The amplitude with intrinsic spin $S=0$ and $L=2$ has the following
form:
\be
A^{M\Lambda}_2(02)&=&(\epsilon^{\Psi M*}\epsilon^{\gamma\Lambda})
X^{(2)}_{\mu\nu}(k_1^\perp)O_{\mu\nu}^{\alpha\beta}
X^{(2)}_{\alpha\beta}(k^\perp) a_{02}(s)\,,
\ee
where $a_{02}(s)\equiv a_{022}(s)$. The amplitude with $SL=12$ has
the form:
\be
A^{M\Lambda}_2(12)&=&\varepsilon_{\mu\alpha\tau
P}\varepsilon_{\alpha\chi\eta P}\epsilon_\chi^{\Psi
M*}\epsilon^{\gamma\Lambda}_\eta
X^{(2)}_{\tau\nu}(k_1^\perp)O_{\mu\nu}^{\alpha\beta}
X^{(2)}_{\alpha\beta}(k^\perp) a_{12}(s)\,.~
\ee
Taking into account that photon polarization vector is orthogonal to $k_1$ and $P$ we obtain:
\be
A^{M\Lambda}_2(12)&=&\frac{3}{2}(\epsilon^{\Psi
M*}k_1^\perp)\epsilon^{\gamma\Lambda}_\mu k^\perp_{1\nu}
O_{\mu\nu}^{\alpha\beta} X^{(2)}_{\alpha\beta}(k^\perp) a_{12}(s)\,.
\ee

\subsection{The factorization of amplitudes}

The convolution of the polarization vectors leads to the tensor
which describes the structure of the vector particle propagator:
\be
\sum_{\lambda}
\epsilon^\lambda_\mu\epsilon^{\lambda*}_\nu=g^\perp_{\mu\nu}=g_{\mu\nu}-\frac{P_\mu
P_\nu}{P^2}\,,
\label{vp}
\ee
where $P$ is the momentum of the vector particle.

The operator $O^{\alpha\beta}_{\mu\nu}$  which describes the
structure of the intermediate tensor state propagator can be written
as:
\be
O^{\alpha\beta}_{\mu\nu}=\sum_{\lambda\lambda'}
\epsilon^\lambda_\mu\epsilon^{\lambda'}_\nu
O^{\alpha\beta}_{\tau\eta}\epsilon^{\lambda*}_\tau\epsilon^{\lambda'*}_{\eta\,\,\,\,.}
\ee
Here $\epsilon^\lambda$ are polarization vectors for the state with
momentum $P$ and we used the property of Eq.(\ref{vp}). Then
\be
O_{\mu\nu}^{\alpha\beta}
X^{(2)}_{\alpha\beta}(k^\perp)=\sum_{\lambda\lambda'}\epsilon^\lambda_\mu\epsilon^{\lambda'}_\nu
O_{\tau\eta}^{\alpha\beta}
\epsilon^{\lambda*}_\tau\epsilon^{\lambda'*}_\eta
X^{(2)}_{\alpha\beta}(k^\perp)=\sum_{\lambda\lambda'}\epsilon^\lambda_\mu\epsilon^{\lambda'}_\nu
\epsilon^{\lambda*}_\tau\epsilon^{\lambda'*}_\eta
X^{(2)}_{\tau\eta}(k^\perp)\,.
\ee
In this case the amplitudes with production of the tensor state can
be written as the sum for the production of two scalar amplitudes:
\be
A^{M\Lambda}_2(SL)&=&\sum_{\lambda\lambda'}A(SL)^{M\Lambda}_{\lambda\lambda'}\times
T_{\lambda\lambda'}a_{SL}(s)\,,
\label{ampl_t}
\ee
where
\be
T_{\lambda\lambda'}=\epsilon^{\lambda*}_\tau\epsilon^{\lambda'*}_\eta
X^{(2)}_{\tau\eta}(k^\perp)
\label{a_x}
\ee
and
\be
A(20)^{M\Lambda}_{\lambda\lambda'}&=&\epsilon^{\Psi
M*}_\mu\epsilon^{\gamma\Lambda}_\nu
\epsilon^{\lambda}_\mu\epsilon^{\lambda'}_\nu a_{20}(s)\,,
\nonumber \\
A(02)^{M\Lambda}_{\lambda\lambda'}&=&(\epsilon^{\Psi
M*}\epsilon^{\gamma\Lambda}) X^{(2)}_{\mu\nu}(k_1^\perp)
\epsilon^{\lambda}_\mu\epsilon^{\lambda'}_\nu  a_{02}(s)\,,\nonumber \\
A(12)^{M\Lambda}_{\lambda\lambda'}&=&\frac{3}{2}(\epsilon^{\Psi
M*}k_1^\perp)\epsilon^{\gamma\Lambda}_\mu \epsilon^{\lambda}_\mu
k^\perp_{1\nu}\epsilon^{\lambda'}_\nu  a_{12}(s)\,\,.
\ee
The amplitudes $A(SL)^{M\Lambda}_{\lambda\lambda'}$ and
$T^{\lambda\lambda'}$ are the scalar amplitudes and can be
calculated in the different coordinate systems.

\subsection{Helicity basis}

The amplitude Eq.(\ref{a_x}) can be calculated in the rest system of
the intermediate state. The polarization vectors in the helicity
basis have the following form:
\be
\epsilon^+&=&\frac{1}{\sqrt 2}(0;-1,-i,0)\,,\nonumber\\
\epsilon^-&=&\frac{1}{\sqrt 2}(0;+1,-i,0)\,,\nonumber\\
\epsilon^0&=&(0;0,0,1)\,\,\,\,.
\ee
Then:
\be
T_{++}&=&\frac 12\left (X^{(2)}_{xx}-X^{(2)}_{yy}-iX^{(2)}_{xy}-iX^{(2)}_{yx}\right )\,, \nonumber \\
T_{+0}&=&-\frac{1}{\sqrt 2}\left (X^{(2)}_{xz}-iX^{(2)}_{yz}\right)=T_{0+}\,, \nonumber\\
T_{00}&=&X^{(2)}_{zz}\,, \nonumber \\
T_{-+}&=&T_{+-}=-\frac 12\left (X^{(2)}_{xx}+X^{(2)}_{yy}\right
)=\frac 12 X^{(2)}_{zz}=\frac 12 T_{00}\,, \nonumber
\\
T_{-0}&=&\frac{1}{\sqrt 2}\left (X^{(2)}_{xz}+iX^{(2)}_{yz}\right)=T_{0-}=-T^*_{0+}\,, \nonumber\\
T_{--}&=&\frac 12\left
(X^{(2)}_{xx}-X^{(2)}_{yy}+iX^{(2)}_{xy}+iX^{(2)}_{yx}\right
)=T^*_{++}\,\,.
\ee
In this system the relative momentum of the final meson can be
written as:
\be
k^\perp&=&|{\bf
k}|(0;\sin\Theta_\pi\cos\phi_\pi,\sin\Theta_\pi\sin\phi_\pi,\cos\Theta_\pi)\,,
\ee
Then
\be
T_{++}&=&\frac 34|{\bf
k}|^2\sin^2\Theta_\pi\,e^{-2i\phi_\pi}=\frac{\sqrt 3|{\bf
k}|^2}{\sqrt 2}d^2_{2,0}(\Theta_\pi) \,e^{-2i\phi_\pi}\,,
\nonumber \\
T_{+0}&=&-\frac{3|{\bf k}|^2}{4\sqrt
2}\sin(2\Theta_\pi)e^{-i\phi_\pi}=\frac{\sqrt 3|{\bf
k}|^2}{2}d^2_{1,0}(\Theta_\pi)e^{-i\phi_\pi},
\nonumber\\
T_{00}&=&\frac 32|{\bf k}|^2\left (\cos^2\Theta_\pi-\frac 13\right)
=|{\bf k}|^2\,d^2_{0,0}(\Theta_\pi)\,\,.
\label{Tt}
\ee

In the rest system of the $\jpsi$ particle the photon vector is
equal to:
\be
k_1=|{\bf
k_1}|(1;\sin\Theta_\gamma\cos\phi_\gamma,\sin\Theta_\gamma\sin\phi_\gamma,\cos\Theta_\gamma)\,.~~
\label{k10}
\ee

Due to the gauge invariance photon has only 2 polarization vectors
$\epsilon^{\gamma\pm}$. If we rotate them to the system where photon
momentum has components from Eq(\ref{k10}), then
\be
\epsilon^{\gamma+}&=\frac{1}{\sqrt 2}(
0;&-\cos\Theta_\gamma\cos\phi_\gamma+i\sin\phi_\gamma,
-\cos\Theta_\gamma\sin\phi_\gamma-i\cos\phi_\gamma,+\sin\Theta_\gamma)\,,\nonumber\\
\epsilon^{\gamma-}&=\frac{1}{\sqrt 2}(
0;&+\cos\Theta_\gamma\cos\phi_\gamma+i\sin\phi_\gamma,
+\cos\Theta_\gamma\sin\phi_\gamma-i\cos\phi_\gamma,
-\sin\Theta_\gamma)\,.~~
\label{eps_gamma}
\ee
The $\jpsi$ also have only two polarization vectors:
\be
\epsilon^{\Psi+*}&=&\frac{1}{\sqrt 2}(0;-1,+i,0)\,,\nonumber\\
\epsilon^{\Psi-*}&=&\frac{1}{\sqrt 2}(0;+1,+i,0)\,.
\ee
Then for the production of the scalar state:
\be
A^{++}_0(00)&=&\frac
12(-\cos\Theta_\gamma\cos\phi_\gamma+i\sin\phi_\gamma
+i\cos\Theta_\gamma\sin\phi_\gamma
-\cos\phi_\gamma)\,a_0(s)\nonumber\\&=&
-\frac{1+\cos\Theta_\gamma}{2}e^{-i\phi_\gamma}\,a_0(s)=-d^1_{1,1}(\Theta_\gamma)\,e^{-i\phi_\gamma}\,a_0(s)\,.
\ee
In the Ablikim et al. \cite{Ablikim:2015umt} paper the corresponding
amplitude is expressed as:
\be
A^{++}_0(E_1)=-d^1_{1,1}(\Theta_\gamma)\,e^{-i\phi_\gamma}\frac{1}{\sqrt
2}\, N_1 N_0 V_{0,1}(s)\,,
\ee
where
\be
N_J=\sqrt{\frac{2J+1}{4\pi}}\,.
\ee

Using the normalization of the amplitude:
\be
\int d\Omega_\gamma d\Omega_\pi |A^{++}_0|^2=\frac 12 |V_{0,1}(s)|^2
\ee
we obtain  the following relation:
\be
a_0(s)=\frac{N_1N_0}{\sqrt
2}\,V_{0,1}(s)=\frac{1}{4\pi}\sqrt{\frac32}\,V_{0,1}(s)\,.
\ee
The polarization vectors $\epsilon^\pm$ related to the tensor
particle produced in the two pion channel coincide with the
polarization vector $\epsilon^{\gamma}$ in its rest system. The zero
component can be calculated by corresponding rotation.
\be
\epsilon^0&=&-\frac{P_0}{\sqrt{s}}(-\frac{|{\bf
P}|}{P_0};\sin\Theta_\gamma\cos\phi_\gamma,
\sin\Theta_\gamma\sin\phi_\gamma,\cos\Theta_\gamma)\,,~~ \nonumber \\
\epsilon^{+}&=&\frac{1}{\sqrt 2}(
0;-\cos\Theta_\gamma\cos\phi_\gamma-i\sin\phi_\gamma,
-\cos\Theta_\gamma\sin\phi_\gamma+i\cos\phi_\gamma,+\sin\Theta_\gamma)\,,\nonumber\\
\epsilon^{-}&=&\frac{1}{\sqrt 2}(
0;+\cos\Theta_\gamma\cos\phi_\gamma-i\sin\phi_\gamma,
+\cos\Theta_\gamma\sin\phi_\gamma+i\cos\phi_\gamma,
-\sin\Theta_\gamma)\,.~~
\ee
It means that
\be
\epsilon^{\gamma+}\epsilon^+=-1\,,\qquad
\epsilon^{\gamma+}\epsilon^-=\epsilon^{\gamma+}\epsilon^0=0
\ee
and we obtain
\be
A(20)^{++}_{++}&=\frac{1}{\sqrt 2}(-\epsilon^+_x+i\epsilon^+_y)
&=-d^1_{1,-1}(\Theta_\gamma)e^{-i\phi_\gamma}\,, \nonumber \\
A(20)^{++}_{-+}&=\frac{1}{\sqrt 2}(-\epsilon^-_x+i\epsilon^-_y)
&=-d^1_{1,1}(\Theta_\gamma)e^{-i\phi_\gamma}\,,
\nonumber \\
A(20)^{++}_{+0}&=\frac{1}{\sqrt 2}(-\epsilon^0_x+i\epsilon^0_y)
&=-d^1_{1,0}(\Theta_\gamma)e^{-i\phi_\gamma}\frac{P_0}{\sqrt s}\,.
\ee
All other combinations are equal to 0.

Thus, using Eqs.(\ref{ampl_t}),(\ref{Tt}) we get the following
formulae:
\be
A_2(20)^{++}\!=\!\left (\frac{\sqrt3|{\bf k}|^2}{\sqrt 2}
B_2\!+\!\frac{\sqrt 3}{2}|{\bf k}|^2\frac{P_0}{\sqrt
s}B_1+\frac{|{\bf k}|^2}{2}B_0\right )a_{20}\,,~~~
\ee
where
\be
B_0&=&-e^{-i\phi_\gamma}d^1_{1,1}(\Theta_\gamma)d^2_{0,0}(\Theta_\pi)\,,\nonumber
\\
B_1&=&-e^{-i\phi_\gamma}d^1_{1,0}(\Theta_\gamma)d^2_{1,0}(\Theta_\pi)\,e^{-i\phi_\pi}\,,\nonumber
\\
B_2&=&-e^{-i\phi_\gamma}d^1_{1,-1}(\Theta_\gamma)d^2_{2,0}(\Theta_\pi)\,e^{-2i\phi_\pi}\,\,.
\ee
These functions are normalized as:
\be
\int d\Omega_\gamma d\Omega_\pi B_i
B^*_j=\delta_{ij}\frac{16\pi^2}{15}\,.
\ee

For the next combination we can reduce $X^{(2)}_{\mu\nu}$ to $\frac
32 k^\perp_{1\mu}k^\perp_{1\nu}$ due to traceless property of the
propagator. Then only polarization vectors with $\epsilon^0$ produce
a non-zero result:
\be
A(02)^{++}&=&\frac{3}{2}|{\bf k_1}|^2\frac{M_{J/\psi}^2}{s}
|{\bf k}|^2\,a_{02}(s)\,B_0\,, \nonumber \\
A(12)^{++}&=&\frac{3\sqrt 3}{4}|{\bf k_1}|^2\frac{M_{J/\psi}}{\sqrt
s} |{\bf k}|^2a_{12}(s)\,B_1\,.
\ee
Then we obtain for the full amplitude:
\be
A_2^{++}\!=\!{\sqrt 6}\tilde a_{20}B_2\!+\!\left (\tilde
a_{12}\!+\!\sqrt 3\tilde a_{20}\frac{P_0}{\sqrt
s}\right)\!B_1\!+\!\left (\tilde a_{20}\!+\tilde a_{02}\right
)\!B_0\,,~~~~
\ee
where
\be
\tilde a_{20}&=&\frac{|{\bf k}|^2}{2}a_{20}(s)\,,\nonumber \\
\tilde a_{12}&=&\frac{3\sqrt 3}{4}|{\bf k_1}|^2|{\bf
k}|^2\frac{M_{J/\psi}}{\sqrt s}a_{12}(s)\,,\nonumber \\
\tilde a_{02}&=&\frac 32\frac{M_{J/\psi}^2}{s}|{\bf k_1}|^2|{\bf
k}|^2a_{02}(s)\,.
\ee

Let us rewrite Eq.(\ref{hel_ampl}) in the terms of $B$-functions:
\be
A_2(E_1)^{++}&=&\frac{N_1N_2}{\sqrt{20}}\left(
\sqrt{6}B_2+\sqrt{3}B_1+B_0\right ) V_{2,1}\,,\nonumber
\\
A_2(M_2)^{++}&=&\frac{N_2N_2}{\sqrt{20}}\left(-\sqrt{2}B_2+B_1+\sqrt{3}B_0\right
)V_{2,2}\,,\nonumber
\\
A_2(E_3)^{++}&=&\frac{N_3N_2}{\sqrt{70}}\left(B_2-\sqrt{8}B_1+\sqrt{6}B_0\right
)V_{2,3}\,.
\ee
These combinations are orthogonal to each another and normalized to
the $\frac 12 |V_{2,j}|^2$. Then in terms of multipoles (see
Eq.(\ref{mult})):
\be
A_2(E1)^{++}&=&\frac{1}{4\pi}\left(
\sqrt{6}B_2+\sqrt{3}B_1+B_0\right )\frac{\sqrt3}{2}E_1\,,\nonumber
\\
A_2(M2)^{++}&=&\frac{1}{4\pi}\left(-\sqrt{2}B_2+B_1+\sqrt{3}B_0\right
)\frac{\sqrt5}{2}M_2\,,\nonumber
\\
A_2(E3)^{++}&=&\frac{1}{4\pi}\left(B_2-\sqrt{8}B_1+\sqrt{6}B_0\right
)\frac{1}{\sqrt{2}}E_{3}\,.
\ee
Then the full amplitude can be written as:
\be
4\pi\,A^{++}_2&=&B_2\frac{1}{\sqrt{2}}\left
(3E_1-\sqrt{5}\,M_2+E_{3}\right )\nonumber \\
&+&B_1\frac{1}{2}\left
(3E_{1}+\sqrt{5}\,M_{2}-4E_{3}\right )\nonumber \\
&+&B_0\frac{1}{2}\left (\sqrt3 E_{1}+\sqrt{15}
M_{2}+\sqrt{12}E_{3}\right )\,.
\ee
And we obtain:
\be
E_{1}&=& \frac{4\pi}{5\sqrt{3}} \left(\tilde {a}_{02}+\sqrt{3}
\tilde {a}_{12}+
\tilde {a}_{20}\left (7+3\frac{P_0}{\sqrt s}\right)\right)\,, \nonumber \\
M_{2}&=& \frac{4\pi}{3\sqrt 5} \left(\sqrt{3} \tilde {a}_{02}+\tilde
{a}_{12}-
\sqrt{3} \tilde {a}_{20}\left(1-\frac{P_0}{\sqrt s}\right)\right)\,,\nonumber \\
E_{3}&=& \frac{8\pi}{15} \left(\sqrt{3} \tilde {a}_{02}-2 \tilde
{a}_{12}+ 2\sqrt{3} \tilde {a}_{20}\left(1-\frac{P_0}{\sqrt
s}\right)\right)\,\,.
\ee
At low energies where $P_0/\sqrt{s}\sim 1$ the amplitude with S-wave
contributed only to the multipole $E_1$. And the reverse equations:
\be
\tilde{a}_{20}&=&\frac{1}{8\pi\sqrt 3} \left(3 \sqrt{5} E_1-\sqrt5 M_2+\sqrt{7}E_3 \right)\,, \nonumber \\
\tilde{a}_{12}&=& \frac{\sqrt{3}}{8\pi\sqrt{s}}\left( \sqrt 3 E_1
\left(\sqrt{s}-P_0\right)-\sqrt 7E_3
\left(P_0+4\sqrt{s}\right)\right .\nonumber \\&+&\left .
\sqrt 5M_2 \left(P_0+\sqrt{s}\right)\right)\,,\nonumber \\
\tilde{a}_{02}&=& \frac{\sqrt 5}{8\pi\sqrt 3} \left(\sqrt{35} E_3+4
M_2\right)\,.
\ee

\section{Conclusion}

We have developed the covariant approach for the partial wave
analysis of the data for the radiative decay of the vector mesons
into two pseudoscalar particles. The approach is based on the $LS$
decomposition of the amplitudes and allowed us in past to perform
the combined analysis of the data on the $\pi\pi$ scattering and the
data on the proton--antiproton annihilation. The obtained equations
which connect the ($SL$) partial wave amplitudes with the amplitudes
extracted in the framework of the helicity approach allowed us to
include in the database the partial wave amplitudes from  the
$\jpsi$ radiative decay extracted in the helicity approach. The
combined analysis showed a prominent signal in the region of the
scalar glueball (see \cite{Sarantsev:2021ein}).

\centerline{\small \bf ACKNOWLEDGE}

The present work was done as a part of the Russian Science
Foundation project (RSF 22-22-00722).

\end{document}